\documentclass[showpacs,amsmath,amssymb,twocolumn,nofootinbib]{revtex4}
\usepackage{epsfig}
\input{epsf}
\usepackage{amssymb}
\begin{document}
\title{Networks of cosmological histories, crossing of the phantom divide line and potentials with cusps}

\author{Francesco Cannata}
\affiliation{Dipartimento di Fisica and INFN, Via 
Irnerio 46,40126 Bologna,Italy}
\author{Alexander Y. Kamenshchik}
\affiliation{Dipartimento di Fisica and INFN, Via 
Irnerio 46,40126 Bologna,Italy\\
L.D. Landau Institute for Theoretical Physics of the 
Russian Academy of Sciences, Kosygin str. 2, 
119334 Moscow, Russia}

\begin{abstract}  
We discuss the phenomenon of the smooth dynamical gravity induced 
crossing of the phantom divide line in a framework of simple cosmological models where it appears to occur rather naturally, provided the potential of the unique scalar field has some kind of cusp. The behavior of cosmological trajectories in the vicinity of the cusp is studied in 
some detail and a simple mechanical analogy is presented. The phenomenon of certain complementarity between
the smoothness of the spacetime geometry and matter equations of motion is elucidated.
We introduce a network of cosmological histories and qualitatively describe some of its properties. 
\end{abstract}
\maketitle 

\section{Introduction}

The discovery of cosmic acceleration \cite{cosmic} has 
stimulated a construction of a class of dark energy models 
\cite{dark,quintessence,darkmodel} describing this effect. 
This dark energy should possess a negative pressure such that the relation
between pressure and energy density $w$ is less than $-1/3$.
Some observations indicate that the present day 
value of the parameter $w < -1$ provides the best fit.  
The corresponding dark energy has been named phantom dark energy \cite{phantom}.
According to some authors, the analysis of observations, 
permits  
to specify the existence of the moment when the universe
changes the value of the parameter $w$ from that the region $w > -1$ to 
$w < -1$ \cite{phant-obs,phant-obs1}. This transition is 
called ``the crossing of the phantom divide line''.  

It is easy to see, that the standard minimally coupled scalar 
field cannot give rise to the phantom dark energy, because in this model
the absolute value of  energy density is always greater than that of 
pressure, i.e. $|w| < 1$. A possible  way out of this situation 
is the consideration of the scalar field models with the negative 
kinetic term.  
 Thus, the important problem arising in connection with 
the phantom energy is  the crossing of the phantom 
divide line. The general belief is that while this crossing is not admissible
in simple minimally coupled models its  explanation  
 requires  more complicated models such as multifield ones
or models with non-minimal coupling between scalar field and gravity 
(see e.g. \cite{divide}).
  
 In our preceding paper \cite{we}
 we have described the phenomenon of the change 
 of sign of the kinetic term of the scalar field implied by  Einstein  
 equations. Now we try to answer the 
 question how general is this phenomenon. In other words, is it some curiosity
 arising due to a very particular choice of the form of a potential 
 of a scalar field and of initial conditions or it is rather a typical  
 phenomenon. In doing  this, we introduce a notion of the network of cosmological histories and qualitatively describe its properties. Admittedly, our research looks  purely academical, but we stress that the persisting signals in favor of the dark energy 
equation of state parameter $w < -1$ \cite{phant-obs1} justifies the interest in this topic and, in particular, in our systematic approach.  The structure of the paper is as follows: Sec. II is devoted to a brief recapitulation of the technique of reconstruction of scalar (phantom) potentials and the transformation of the unique scalar-phantom field (i.e. the crossing of the phantom divide line; in Sec. III we discuss the role of initial conditions for the phantom divide line crossing phenomenon; Sec. IV is devoted to the analysis of a simple mechanical system with the potential with a cusp, which is strictly motivated by the requirement of the possibility of crossing of the phantom divide line within a framework which allows a unique scalar field; in Sec. V we introduce the notion of network of cosmological histories; the section VI contains concluding remarks 
In the Appendix A  we shall present simple solvable examples of 
 such networks. In the Appendix B we shall describe 
 some potentials as functions of time or scalar field, for which exact solution of the Einstein equations is possible.
 
 \section{Reconstruction of scalar (phantom) potentials and dynamical crossing of the phantom divide line}
 Let us remember the main points of our approach.
 We have considered the minimally coupled scalar 
 field with an exponential potential  
 \begin{equation}
V_0(\phi) \sim \exp\left(-\frac{3\sqrt{1+w}\phi}{2}\right).
\label{exponent}
\end{equation}
Such potentials are widely studied in cosmology 
\cite{exp} and have a particular solution corresponding to 
a special choice of initial conditions, which implies
the power-law 
behavior of the cosmological radius
\begin{equation}
a(t) = a_0 t^{\frac{2}{3(1+w)}},
\label{radius}
\end{equation}
or, in other terms,
\begin{equation}
h(t) = \frac{2}{3(1+w)t},
\label{Hubble}
\end{equation}
where $h=\dot{a}/a$ is the Hubble variable. 
Changing the initial conditions (i.e. the value of $\dot{\phi}$ corresponding to some value of $\phi$)  one shall have 
other laws of cosmological evolution, which cannot 
be presented explicitly, but which have the same 
qualitative behavior as (\ref{radius}).
 
However, one can consider the potential (\ref{exponent} not as a function of the scalar field $\phi$, but as a function of the cosmic time parameter 
$t$ as was done in \cite{Chervon,Yurov,Yurov1,we}.
In this case the Einstein (Friedmann) equation 
is equivalent to the linear second-order differential 
equation   
\begin{equation}
\ddot{\psi} = 9V(t)\psi,
\label{equation}
\end{equation}
for the volume function 
\begin{equation}
\psi(t) \equiv a^{3}(t).
\label{volume} 
\end{equation}
The potential $V(t)$ corresponding to the exponential potential (\ref{exponent}) and 
to the choice of the initial conditions providing the 
evolution (\ref{radius}) is 
\begin{equation}
V_1(t) = \frac{2(1-w)}{9(1+w)^2t^2}.
\label{pot-time}
\end{equation}
Equation (\ref{equation}) with the potential (\ref{pot-time}) has a {\it general} solution 
\begin{eqnarray}
&&\psi(t) = \psi_1 t^{\alpha_1} + \psi_2 t^{\alpha_2},\nonumber \\
&&\alpha_1 = \frac{2}{1+w},\nonumber \\
&&\alpha_2 = \frac{w-1}{1+w},
\label{solution}
\end{eqnarray}
where $\psi_1$ and $\psi_2$ are nonnegative constants.  If $\psi_2 = 0$, one reproduces the 
cosmological evolution (\ref{radius}), while if 
$\psi_1 = 0$ the cosmological evolution requires 
the presence of the phantom scalar field with the 
negative sign of the kinetic term. The most interesting situation reveals when, both the coefficients $\psi_1$ and $\psi_2$ are nonzero. 
Simple calculation shows that in this case the Hubble variable has a maximum point where, the kinetic term is forced to change its sign.   
\footnote{Notice that in papers \cite{reconstruct} the technique of the reconstruction of the potentials as functions of the redshift parameter was developed. 
The reconstruction approach is (to a large extent) model
independent. So, it may show that the phantom boundary is crossed in nature,
but do not specify a class of potentials as functions of $\phi$ in which this can
be described in a consistent way}.

 Now, fixing the relation between these coefficients 
one can ask what is the corresponding form of the 
potential as a function of the scalar field $\phi$. 
The explicit form of this function cannot be found,
but we have studied \cite{we} its asymptotic behavior
at singularities and in the point $t_0$, where the kinetic 
term changes the sign and  (de)-phantomization occurs, or, in other words, where the universe crosses the phantom divide line. In the neighborhood      
of the moment $t_0$, which depend on the relation
between the coefficients $\psi_1$ and $\psi_2$ 
\cite{we}, the potential has the form
\begin{eqnarray}
&&V(\phi) = 
\frac{2(1-w)}{9(1+w)^2\left(t_0+
\left(\sqrt{\frac{27}{8H}}\phi\right)^{2/3}\right)^2},\nonumber \\
&&H =  \frac{\sqrt{8(1-w)}(3-w)^2\psi_1\psi_2t_0^{-\frac{4w}{1+w}}}
{(1+w)^3\left(\psi_1 t_0^{\frac{2}{1+w}} + \psi_2 t_0^{\frac{w-1}{w+1}}
\right)^2}.
\label{pot-zero}
\end{eqnarray}
The scalar field behaves correspondingly as 
\begin{equation}
\phi \sim (t-t_0)^{3/2}.
\label{scal}
\end{equation} 

\section{Initial conditions and phantom divide line crossing}
In the previous section  we have implicitly chosen such initial conditions so that simultaneously the scalar field and 
its time derivatives vanish at the point $t_0$, which 
in turn implies the crossing of the phantom divide line. However, let us fix the form of the potential 
as a function of the scalar field $\phi$ and see how 
the behavior of the universe will change depending 
on the initial conditions. 
Suppose that when the cosmic 
parameter is close to $t_0$ the potential behaves as 
\begin{equation}
V(\phi) = \frac{1}{(A + B\phi^{2/3)^2}}.
\label{pot-close} 
\end{equation}
The scalar field can have the following form:
\begin{equation}
\phi(t) = \phi_0 + \phi_1(t-t_0)^{\alpha},
\label{scal-close}
\end{equation}
where $\alpha > 1$ to provide the vanishing of the time derivative of the scalar field at the point $t_0$. The situation described above
and corresponding to the (de) - phantomization of 
the scalar field \cite{we} 
realizes when $\phi_0 = 0$ and $\alpha = 3/2$. 
Let us consider the cases when $\phi_0 \neq 0$,
analyzing the Klein-Gordon equation 
\begin{equation}
\ddot{\phi} + 3h \dot{\phi} +\frac{dV}{d\phi} = 0.
\label{KG}
\end{equation}
The terms here behave as 
\begin{equation}
\ddot{\phi} = \alpha(\alpha-1)\phi_1(t-t_0)^{\alpha-2},
\label{sec-der}
\end{equation}
\begin{equation}
\dot{\phi} = \alpha\phi_1(t-t_0)^{\alpha-1},
\label{first-der}
\end{equation}
\begin{equation}
\frac{dV}{d\phi} = -\frac{4B}{3\phi^{1/3}(A+B\phi^{2/3})^3},
\label{pot-der}
\end{equation}
\begin{equation}
h = \sqrt{\frac12\alpha^2\phi_1^2(t-t_0)^{2\alpha-2} +
\frac{1}{(A+B\phi_0^{2/3})^2}}.
\label{Hubble-behave}
\end{equation} 

Let us consider the case $\alpha > 2$. 
In this case the first and the second terms of Eq. (\ref{KG}) vanish at $t \rightarrow 0$ while the term 
(\ref{pot-der}) is a constant. Thus, we have come to
a contradiction and this case should be disregarded.

Now, considering the case $\alpha = 2$, we see that
while the term containing the first derivative of 
the scalar field vanishes the term with the second derivative becomes a constant ($2\phi_1$) to be 
equated to the the term (\ref{pot-der}) up to the sign. 
Therefore, we obtain the relation 
\begin{equation}
\phi_1 = \frac{2B}{3\phi^{1/3}(A+B\phi^{2/3})^3}.
\label{relation}   
\end{equation}
It is easy to see that in this case 
the time derivative of the Hubble parameter $h$ 
behaves as 
\begin{equation}
\dot{h} \sim (t-t)^2
\label{Hubble-der}
\end{equation}
in contrast to the behavior $\dot{h} \sim t-t_0$,
corresponding to the case $\phi_0 = 0$, 
considered in \cite{we}. Apparently, the behavior 
(\ref{Hubble-der}) means that the Hubble variable 
has at the moment $t_0$ an inflection point instead 
of the extremum point and hence, the kinetic term 
for the scalar field conserves its sign. 

In the case when $1 < \alpha < 2$ the second derivative term diverges while other terms are finite,
so this behavior cannot be realized.   

Until now we have considered the cosmological trajectories having at some moment 
the vanishing time derivative $\dot{\phi}$ of the scalar field, while $\phi$ could have zero or nonzero values.  However, there is also another class of  trajectories characterized by the fact that at some moment $t_1$ the scalar field vanishes, while its first time derivative is different from zero. Such trajectories are characterized by the equation:
\begin{equation}
\phi(t) = v_1(t-t_1) + v_{5/3}(t-t_1)^{5/3},
\label{singul}
\end{equation}
where the constants $v_1$ and $v_{5/3}$ satisfy the condition
\begin{equation}
v_{5/3} = -\frac{6B}{5A^3 v_1^{1/3}},
\label{condsing}
\end{equation}
which can be easily obtained from Eqs. (\ref{KG}) and (\ref{pot-der}). 
Calculating the second time derivative of the Hubble variable, one can see that 
it behaves as $\ddot{h}(t) \sim (t-t_1)^{-1/3}$ that means that higher curvature invariants including the curvature derivatives diverge and one encounters the so called 
soft singularity. The problem of removal of such singularities is of interest by intself, but will not be considered here. 

Thus, let us summarize the results of the above consideration. We have found three types of the trajectories. For the first   family of the 
trajectories the time derivative of the scalar
field vanishes at the moment $t_0$. This family is a 
one-parameter one and is parametrized by the value 
of the scalar field $\phi_0$, which should be different from zero at the moment $t_0$. 
All the nonzero values of $\phi_0$ require $\alpha = 2$, the value of $\phi_1$ is fixed by relation (\ref
{relation}), the Hubble variable has an inflection point 
and the kinetic term conserves its sign.

The second family contains the only trajectory with 
$\phi_0 = 0$. Now  the term (\ref{pot-der}) 
becomes singular and behaves as $(t-t_0)^{-\alpha/3}$, the term with the first derivative vanishes, while 
the second derivative term behaves like $(t-t_0)^{\alpha-2}$. Consistency requires $\alpha = 3/2$ 
and the value of $\phi_1$ is uniquely fixed. The Hubble variable has at the moment $t_0$ an extremum and crossing of the phantom divide line occurs.  
The third family is characterized by Eqs. (\ref{singul}), (\ref{condsing}) and suffers from  a soft cosmological singularity.

 It seems 
that in some sense one can consider the possibility 
of the crossing of the phantom divide line as was 
described in \cite{we} as  exceptional and the corresponding initial conditions as having measure zero. Nevertheless, we would like to show that 
considering broader set of potentials and initial 
conditions, one can come to the  conclusion, that 
conditions providing the crossing are, roughly speaking, commensurable to  those excluding it.  We postpone the corresponding consideration until the fifth section, 
while the next section will be devoted to an analysis of the simple mechanical model 
of the particle moving in the potential with a cusp.

\section{Mechanical analog: a particle moving in a potential with a cusp} 
Let us consider a one-dimensional problem of a classical point particle moving in the potential
\begin{equation}
V(x) = \frac{V_0}{(1+x^{2/3})^2},
\label{classical}
\end{equation}
where $V_0 > 0$.
The equation of motion is 
\begin{equation}
\ddot{x} -\frac{4V_0}{3(1+x^{2/3})^3 x^{1/3}} = 0.
\label{classical1}
\end{equation}
We consider three classes of possible motions characterized by the value of the energy
$E$. The first class consists of the motions when $E < V_0$. Apparently, the particle 
with $x < 0, \dot{x} > 0$ or with $x > 0, \dot{x} < 0$ cannot reach the point $x = 0$ 
and stops at the points $\mp \left(\sqrt{\frac{V_0}{E}}-1)\right)^{3/2}$ respectively.
This class of motions corresponds to the class of cosmological evolutions described in 
the preceding chapter when the Hubble variable has an inflection point and the universe 
does not cross the phantom divide line. 

The second class includes the trajectories when $E > V_0$. In this case the particle crosses the point $x = 0$ with nonvanishing velocity and this case correspond to the cosmological evolution given by Eq. (\ref{singul}), but naturally there is nothing singular in it, because in our simple mechanical problem, we do not have nothing similar 
to the Hubble variable and the spacetime geometry defined by its behavior. 

If we have a fine tuning such that $E = V_0$, we encounter an exceptional case. 
Now the  trajectory satisfying Eq. (\ref{classical1})in the vicinity of the point $x = 0$ can behave as 
\begin{equation}
x = C(t_0-t)^{3/2},
\label{classical2}
\end{equation}
where 
\begin{equation}
C = \pm \left(\frac{16V_0}{9}\right)^{3/4}
\label{C-define}
\end{equation}
and $t \leq t_0$. 
It is easy to see that independently of the sign of $C$ in Eq. (\ref{C-define}) 
the signs of the particle coordinate $x$ and of its velocity $\dot{x}$ are opposite and hence, the particle can arrive in finite time to the point of the cusp of the potential 
$x = 0$. 

Another solution reads as 
\begin{equation}
x = C(t-t_0)^{3/2},
\label{classical4}
\end{equation}
where $t \geq t_0$.
This solution describes the particle going away from the point $x = 0$. 
Thus, we can combine the branches of the solutions (\ref{classical2}) and (\ref{classical4}) 
in four different manners and there is no way to choose if the particle arriving to the point 
$x=0$ should go back or should pass the cusp of the potential (\ref{classical}). It can stop at the top as well. Such a ``degenerate'' behavior of the particle in this third case is connected 
with the fact that  this trajectory is the separatrix between two one-parameter families described above. 
At the moment there is not yet any strict analogy between this separatrix and the cosmological evolution describing the phantom divide line. In order to establish a closer analogy  and to understand what is the crucial difference between mechanical consideration and  general relativistic one, we 
can try to introduce a friction term into the Newton equation (\ref{classical1})
\begin{equation}
\ddot{x} + \gamma\dot{x}-\frac{4V_0}{3(1+x^{2/3})^3 x^{1/3}} = 0.
\label{classical5}
\end{equation} 
It is easy to check that if the friction coefficient $\gamma$ is a constant one does not have 
a qualitative change in respect to the discussion above. Let us asuume for $\gamma$ the dependence 
\begin{equation}
\gamma = 3\sqrt{\frac{\dot{x}^2}{2}+V(x)}.
\label{gamma} 
\end{equation}
then
\begin{equation}
\dot{\gamma} = -\frac{3}{2}\dot{x}^2
\label{gammadot}
\end{equation}
and 
\begin{equation}
\ddot{\gamma} = -3\ddot{x}\dot{x}.
\label{gammaddot}
\end{equation}
The trajectory arriving to the cusp with vanishing velocity is still described by 
the solution (\ref{classical2}). Consider the particle coming to the cusp from the left 
($C < 0$. It is easy to see that the value of $\dot{gamma}$ at the moment $t_0$ tends to zero,
while its second derivative $\ddot{\gamma}$ given by Eq. (\ref{gammaddot}) is 
\begin{equation}
\ddot{\gamma}(t_0) = \frac98 C^2 > 0.
\label{gammaddot1}
\end{equation}
Thus, it looks like  the friction coefficient $\gamma$ reaches its minimum value at $t = t_0$. 
Let us suppose now that the particle is coming back to the left from the cusp and its motion is described by Eq. (\ref{classical4}) with negative $C$. A simple check shows that 
in this case 
\begin{equation}
\ddot{\gamma}(t_0) = -\frac98 C^2 < 0.
\label{gammaddot2}
\end{equation}
Thus, from the point of view of the subsequent evolution this point looks as a maximum 
for the function $\gamma(t)$. In fact, it means simply that the second derivative of 
the friction coefficient has a jump at the point $t = t_0$.
It is easy to check that if instead of choosing the motion to the left, we shall move forward our particle to the right from the cusp ($C>0$), the sign of $\ddot{gamma}(t_0)$ remains negative as in Eq. (\ref{gammaddot2}) and hence we have the jump of this second derivative again. If one would like to avoid 
this jump, one should try to change the sign in Eq. (\ref{gammaddot}). 
To implement it in a self-consistent way one can substitute Eq. (\ref{gamma}) by 
\begin{equation}
\gamma = 3\sqrt{-\frac{\dot{x}^2}{2}+V(x)}
\label{gamma1} 
\end{equation}
and Eq. (\ref{classical5}) by 
\begin{equation}
\ddot{x} + \gamma\dot{x}+\frac{4V_0}{3(1+x^{2/3})^3 x^{1/3}} = 0.
\label{classical51}
\end{equation} 
In fact, it is exactly that what 
happens automatically in cosmology, when we change the sign of the kinetic energy term for the scalar field, 
crossing the phantom divide line. Naturally, in cosmology the role of $\gamma$ is played by the Hubble variable $h$.  The jump of the second derivative of the friction coefficient $\gamma$ corresponds to the divergence of the third time derivative of the Hubble variable, which represents some kind of soft cosmological singularity (this singularity is even softer than that considered in the preceding section
for the family of cosmologies characterized by Eq. (\ref{singul}) where already the second time derivative of the Hubble variable was divergent). 

Thus, one seems to confront the problem of choosing between two alternatives: 1) to encounter a weak singularity in the spacetime geometry; 2) to change the sign of the kinetic term for matter field.  
In this paper  we pursue the second alternative insofar as 
we privilege the smoothness of spacetime geometry and consider equations of motion for matter as less 
fundamental than the Einstein equations (see also more detailed discussion in \cite{we}).

\section{Potential as a cosmic time function and the network of cosmological histories}
Indeed, all the previous considerations were based on the treatment of the potential as a fixed 
function of the scalar field and changing initial conditions for this scalar field. As was already 
emphasized \cite{Chervon,Yurov,Yurov1} one can use an alternative treatment of the potential $V$     as a function of time. In the example considered 
in \cite{we}, it was shown that for $V(t)$ instead 
the initial conditions which exclude the crossing of the phantom divide line are exceptional. 

Let us try 
to imagine a set of possible cosmologies as a two-dimensional surface (see the picture below) where 
the vertical lines represent the potentials as functions of $\phi$ while the horizontal lines represent the potentials as functions of $t$.      
\begin{picture}(200,200)(0,0)
\put(40,40){\line(0,1){150}}
\put(170,40){\line(0,1){150}}
\put(20,60){\line(1,0){170}}
\put(20,160){\line(1,0){170}}
\put(170,60){\circle*{5}}
\put(180,50){A}
\put(170,160){\circle{10}}
\put(180,150){D}
\put(40,60){\circle*{5}}
\put(50,50){B}
\put(40,160){\circle*{5}}
\put(50,150){C}
\put(170,100){$V_0(\phi)$}
\put(90,50){$V_1(t)$}
\put(100,165){$V_3(t)$}
\put(40,100){$V_2(\phi)$}
\put(25,20){\makebox(0,0)[l]{\bf Network of cosmological histories}}
\end{picture}

We shall call it network of cosmological histories. 
Moving along these lines corresponds to changing 
of initial conditions keeping fixed the form of the potential. For example, for the vertical line corresponding to the exponential potential (\ref{exponent}) (see the line $V_0(\phi)$ in the picture, the black circle  $A$ 
corresponds to the power-law cosmological evolution given by formulae (\ref{radius} ) and (\ref{Hubble}))
there are no initial 
conditions allowing the crossing of the phantom divide line. Now, fixing the initial conditions providing the cosmological evolution (\ref{radius}), 
(\ref{Hubble}) we obtain the potential $V_1(t) \sim 
1/t^2$ and moving along the corresponding horizontal line one finds that at all the points, excluding the point of intersection $A$, there is the (de)-phantomization effect. Let us stop now at some point 
of the horizontal line (say, $B$) and fix the corresponding form 
of the potential $V_2(\phi)$ as a function of $\phi$. Then, one can begin moving along the vertical line, corresponding to this form of 
$V_2(\phi)$. The intersection point $B$ corresponds to such choice of initial conditions when $\phi$ and $\dot{phi}$ vanish 
simultaneously at some moment $t_0$. 
And this is the only point of the vertical line, where the crossing of the phantom divide line occurs as was shown above. To come to 
this conclusion we have used only some general 
properties of the form of the potential around a 
candidate moment of the crossing of the phantom divide line. 

Now, we undertake a similar analysis showing that 
for a horizontal line (potential as a function of $t$) 
almost all the initial conditions imply the crossing effect. 
Let us consider a potential $V_3(t)$ corresponding to the horizontal line intersecting the vertical line $V_2(\phi)$ at the point $C$. 
We shall look for a cosmological evolution 
where the Hubble variable has an extremum at the 
moment $t_0$. A regular potential can be represented around this moment as 
\begin{equation}
V(t) = \frac19(U_0 + U_1(t-t_0)).
\label{pot-gen} 
\end{equation}   
Then Eq. (\ref{equation}) acquires the form
\begin{equation}
\frac{d^2\psi}{d\tau^2} =(U_0 + U_1\tau)\psi,
\label{equation1}
\end{equation} 
where $\tau \equiv t-t_0$. 
We shall look for two independent solutions in the vicinity of $\tau =0$;
in the form 
\begin{equation} 
\psi(\tau) = \sum_{n=0}^{\infty} a_n \tau^n.
\label{expansion} 
\end{equation} 
The recurrent relation between the coefficients
$a_n$ is 
\begin{equation}
(n+2)(n+1)a_{n+2} = U_0 a_n + U_1a_{n-1}.
\label{recurrent}
\end{equation} 
We define the first independent solution by fixing of the first two coefficients as $a_0=1, a_1=1$ while for the second solution we choose $a_0=1,a_1=0$
(one can easily check that these solutions are independent, calculating the first terms of their Wronskian). 
The leading terms of the general solution of Eq. (\ref{equation1}) has the following form:
\begin{eqnarray}
&&\psi(\tau) = \psi_1\left(1+\tau+\frac{U_0 \tau^2}{2} +
\frac{(U_0+U_1)\tau^3}{6}\right)\nonumber \\
&&+\psi_2\left(1+\frac{U_0 \tau^2}{2} +
\frac{U_1\tau^3}{6}\right),
\label{solution1} 
\end{eqnarray} 
where $\psi_1$ and $\psi_2$ are nonnegative. 
We shall need also the expressions for the first and second time derivatives of the solution (\ref{solution1}):
\begin{equation}
\dot{\psi} = \psi_1\left(1+V_0\tau +\frac{(V_0+V_1)\tau^2}{2}\right) + 
\psi_2\left(V_0\tau +\frac{V_1\tau^2}{2}\right),
\label{first-der1}
\end{equation} 
\begin{equation}
\ddot{\psi} = \psi_1(U_0 + (U_0+U_1)\tau) 
+ \psi_2(U_0 + U_1 \tau).
\label{second-der}
\end{equation} 

The time derivative of the Hubble variable has the form
\begin{equation}
\dot{h} = \frac13 \frac{\ddot\psi\psi - \dot{\psi}^2}
{\psi^2}.
\label{Hubble-structure}
\end{equation} 
It is enough to study the numerator of expression 
(\ref{Hubble-structure}) 
\begin{eqnarray} 
&&\ddot{\psi}\psi -\dot{\psi}^2 = 
(\psi_1+\psi_2)^2U_0 - \psi_1^2 \nonumber \\
&&+(\psi_1+\psi_2)U_1 \tau.
\label{Hubble-structure1}
\end{eqnarray}

To have at the moment $\tau  = 0$ the extremum of the function $h(\tau)$ its time derivative should behave as 
$\dot{h} \sim \tau$, that means that Eq. (\ref{Hubble-structure1}) implies that 
\begin{equation}
(\psi_1+\psi_2)^2U_0 = \psi_1^2,
\label{cond-extr}
\end{equation}   
and 
\begin{equation}
U_1 \neq 0.  
\label{cond-extr1}
\end{equation}     

The condition (\ref{cond-extr}) determines the relation between the coefficients $\psi_1$ and 
$\psi_2$ in terms of the coefficient $U_0$. 
Thus, for a potential $V(t)$ fixing the moment 
of time $t_0$ we fix the constants $U_0$ and 
$U_1$. If the condition (\ref{cond-extr1}) is 
satisfied, choosing the constants $\psi_1$ and 
$\psi_2$ satisfying condition (\ref{cond-extr}) one 
obtains a cosmological evolution undergoing the 
crossing of the phantom divide line.   

\section{Concluding remarks}
Thus, we have seen that considering 
the potential as a function of the scalar field having a cusp  
we should choose exceptional initial conditions
to describe a smooth cosmological evolution undergoing 
the phantom divide line crossing. We have seen also that there is some kind of complementarity between
the smoothness of the spacetime geometry and that of the structure of matter action.

On the hand, considering the potential as a function of time we 
encounter just the following situation: initial conditions implying the crossing of the phantom divide line are typical. That means that roughly speaking in the set of all possible cosmological trajectories those crossing and non-crossing the phantom divide line have measures of the same order. We are not able to construct a rigorous description of this set of cosmological histories, but 
the treatment undertaken above permits to 
make some qualitative remarks. First of all let us notice that a network of vertical lines representing 
the potential as a function of the scalar field and 
horizontal lines , representing potentials as functions
of the cosmic time, is not closed. Namely, let us leave a vertical line at the point $A$ and make some walk along the horizontal line to the point $B$. 
At the point $B$ we shall have a cosmology with the crossing. Then let us make some walk along the vertical line from the point $B$ to 
some point $C$. The cosmology corresponding 
to the point $C$ is free from crossing. Now, traveling 
along the horizontal line from point $C$ we shall always have trajectories with crossing and, hence,  
we shall not have an
opportunity to return to the initial vertical line, which 
should have cosmologies without crossing. (The empty circle $D$ in our picture signifies the absence of  intersection between the 
respective vertical and horizontal lines). 

There is another interesting question. It is well known that for the exponential potential (\ref{exponent}) all the initial conditions imply the cosmologies free from the crossing phenomenon. 
The explanation of this fact is a very simple one:
the effect of smooth crossing \cite{we} can have place only for potentials which at some point 
have a divergent derivative with respect to the scalar field, while the derivative of the exponential potential 
is always regular.

In conclusion we would like to say that the above investigation shows 
that the smooth crossing of the phantom divide line  in simple 
cosmological models could be considered as a rather  natural phenomenon more than an exotic anomaly. The network of cosmological histories constructed for the illustration of this fact seems 
to be interesting object on its own. The points of this  network  represent functions $h(t)$ on which two transformations generated  by means of  
scalar field potentials, treated as functions of $\phi$ or of $t$. Graphically they were represented as displacements along vertical and horizontal lines.  
The network discussed in this paper was constructed  beginning from the simple cosmological evolution 
(\ref{Hubble}). Notice that our initial point of network 
(\ref{Hubble}) is characterized by some value of the equation of state  parameter $w$ and is nothing but a power-law evolution (\ref{radius}). The points 
lying on a horizontal line intersecting this initial power-law point (the point A on  our picture), are described explicitly \cite{we} and have more complicated structure. As far as other points of the network  are concerned we have studied above some of their local properties, but we are not able 
to describe their global topology. It seems unlikely that two different power-law points belong to the same network. If this statement is confirmed one has 
got a family of networks characterized by different values of the initial constant equation of state parameter $w$. In the Appendix A we give an explicit illustration of the simplest network containing an exponential expansion, while in the Appendix B we give some other examples of potentials for which it is possible to find exact solutions.

\section*{Acknowledgements}
We are grateful to A.A. Starobinsky for  fruitful discussions and for  critical reading of the manuscript.
We are grateful to A.A. Andrianov for discussions at the initial stage of this work. 
A.K. was partially supported by RFBR, grant No. 05-02-17450.

\section*{Appendix A. Examples of simple networks of cosmological histories}
We present here a particularly simple and in some way degenerate example of the network of cosmic 
histories which is discrete and contains only three points. 
Let us consider de Sitter universe with 
\begin{equation}
h(t) = H_0.
\label{Hubble-dS}
\end{equation}
The corresponding potential as is well-known 
and as can be easily obtained from the equation 
(see e.g. \cite{we} and references therein)
\begin{equation}
V = \frac{\dot{h}}{3} + h^2,
\label{hdot-pot}
\end{equation}
has the form
\begin{equation}
V(\phi) = H_0^2,
\label{pot-dS}
\end{equation}
while the initial condition is     
\begin{equation}
\dot{\phi} = 0.
\label{initial-dS}
\end{equation}
This initial condition can be defined at any moment of time and the evolution shall conserve it, because
the Klein-Gordon equation (\ref{KG}) for the constant potential   
(\ref{pot-dS}) has the form
\begin{equation}
\ddot{\phi} +3h\dot{\phi} = 0.
\label{KG1}
\end{equation}
Notice that in this case the Klein-Gordon equation has the same form for usual scalar and phantom.
Now, treating the potential as a constant function of time we can easily resolve for this case Eq. (\ref{equation}): 
\begin{equation}
\psi = \psi_1 \exp(3H_0 t) + \psi_2\exp(-3H_0 t).
\label{volum-sol}
\end{equation}
The Hubble function correspondingly is
\begin{equation}
h(t) = H_0 \frac{\psi_1 \exp(3H_0 t) - \psi_2\exp(-3H_0 t)}{\psi_1 \exp(3H_0 t) + \psi_2\exp(-3H_0 t)},
\label{Hubble-degen}
\end{equation}
while its time derivative is 
\begin{equation}
\dot{h}(t) = \frac{12H_0^2 \psi_1\psi_2}{(\psi_1 \exp(3H_0 t) + \psi_2\exp(-3H_0 t))^2}.
\label{Hubble-dot-deg}
\end{equation}
As usual the constants $\psi_1$ and $\psi_2$ are 
nonnegative. Correspondingly we have three opportunities. If $\psi_2 = 0$ this is our initial 
point describing infinitely expanding flat de Sitter universe. If $\psi_1 = 0$, one has an infinitely contracting flat de Sitter universe. If both the coefficients are positive, the Hubble variable is growing and this situation could be realized only 
in the universe where the scalar field has a negative kinetic term (phantom). It is easy to understand that 
a change of the relation $\psi_2/\psi_1$ is simply equivalent to the shift of the cosmic time variable  $t$.  Thus, we have only one cosmic history described by phantom field with the constant potential. 
In this history the universe begins its evolution 
at $t = -\infty$ with a quasi-de Sitter contraction.
It arrives to a minimum possible value of the cosmological radius at some moment $t_{min}$.
This moment can be easily found from Eq. (\ref{Hubble-degen}) however,  its value does not have a particular meaning just like the relation     
$\psi_2/\psi_1$. After the moment $t_{min}$ begins an expansion which at $t \rightarrow \infty$ becomes quasi-de Sitter. Making a convenient shift  of the time variable we can write down Eq. (\ref{Hubble-degen}) in a specially simple form:
\begin{equation}
h(t) = H_0 \tanh 3H_0 t.
\label{tanh}  
\end{equation}
Thus, instead of a horizontal line we obtain 
two discrete points. 

Now, let us treat the constant potential  (\ref{pot-dS}) as a function of $\phi$ changing the initial condition 
for the time derivative of this field. It is easy to show 
that for $\dot{\phi} \neq 0$ there two cosmological 
histories given by the following expression for the Hubble variable:
\begin{equation}
h(t) = \pm H_0 \coth 3H_0 t.
\label{sing}
\end{equation}
These solutions describe a universe which begins 
its evolution from cosmological singularity at 
$t = 0$ and then is infinitely expanding arriving to 
quasi-de Sitter phase at $t \rightarrow \infty$ and a universe which begins its evolution at $t =-\infty$ 
and then contracts arriving to the Big Crunch type 
cosmological singularity at $t = 0$. It is important to notice that the choice of the initial value of the time derivative of the scalar field is not important, because in the process of evolution this value 
runs between $0$ and $\pm \infty$. The sign of this time derivative is not important either because 
the potential of this field is constant. 

Thus, we have presented a schematic example of an exactly solvable cosmological network, which contains only five discrete points.  Two of these points 
represent an infinitely expanding and infinitely contracting de Sitter universes, two of them describe 
universe evolving from the cosmological singularity to an infinite expansion or vice versa, driven by 
a scalar field with constant potential. The last point represents a universe which passes from contraction to expansion and is driven by a phantom field. All the network is characterized by an absolute value of the constant $H_0$. Indeed, beginning from $-H_0$ we arrive to the same network, which was constructed from $H_0$. 

 As a last remark we can give two examples of even more trivial networks of cosmological histories. First of them contains only one point. If one has a negative 
 constant potential $V = -H_0^2$, than the only cosmological history present in the network is that given by $h(t) = -H_0 \tan 3 H_0 t$ describing the evolution from the Big Bang at $t = -\pi/6H_0$ to 
 the Big Crunch at $t = \pi/6H_0$ passing through 
 the point of maximal expansion at $t = 0$.  
 The second network consists of three points. Considering massless scalar field with vanishing potential, one can see that there are three opportunities: if the time derivative of the scalar field is nonzero, then one has either a universe which begins its evolution  from the Big Bang singularity and expands infinitely or a universe which contracts 
 finishing in the Big Crunch singularity. If the time derivative of the scalar field is zero, one has  a static 
 Minkowski  universe. 
  
\section*{Appendix B. Some solvable potentials}  
The concept of the network of cosmological histories 
introduced in this paper is based on the simultaneous use of the notion of potential as function of time and as function of the scalar field. Unfortunately, it is much more difficult to find exactly solvable examples for potentials as functions of scalar fields than for those, treated as functions of time. Thus, we, first give another example of the potential as function of $\phi$ (see, for additional details \cite{we-tach}). 

Let us consider the cosmological evolution given by the Hubble function 
\begin{equation}
h(t) = H_0 \coth \frac{3H_0(1+w)t}{2},
\label{Hubblecom}
\end{equation}
while the cosmological radius behaves as 
\begin{equation}
a(t) = a_0\left(\sinh \frac{3H_0(1+w)t}{2}\right)^
{2/[3(1+w)]}.
\label{radiuscom}     
\end{equation}  
This evolution occurs in the universe filled with 
two perfect fluids: one of them is a cosmological constant and the other is a barotropic fluid with the equation of state parameter $w$. The scalar field 
potential realizing this evolution has the form 
\cite{we-tach}:
\begin{equation}
V(\phi) = H_0^2\left(1+\frac{1-w}{2}\sinh^2\frac{3\sqrt{1+w}\phi}{2}\right).
\label{pot-com}
\end{equation}   
The potential as a function of time is 
\begin{equation}
V(t) = H_0^2 + \frac{(1-w)H_0^2}{2\sinh^2 \frac{3H_0(1+w)t}{2}}.
\label{pot-time-com}
\end{equation}
The general solution of Eq. (\ref{equation}) with potential (\ref{pot-time-com}) can be written as 
usual as 
\begin{eqnarray}
&&\psi(t) = \psi_1\left(\sinh \frac{3H_0(1+w)t}{2}\right)^
{2/(1+w)} \nonumber \\ 
&&+ \psi_2      
\left(\sinh \frac{3H_0(1+w)t}{2}\right)^
{2/(1+w)}\nonumber \\ 
&&\times\int dt' \left(\sinh \frac{3H_0(1+w)'}{2}\right)^{-4/(1+w)}.
\label{solution-com}   
\end{eqnarray}  

The explicit form of the solution (\ref{solution}) is very simple for some special values of the parameter 
$w$. We shall consider the case $w = 0$. Now, the solution is 
\begin{equation}
\psi(t) = \psi_1\sinh^2\frac{3H_0t}{2}
+\psi_2\left(\coth\frac{3H_0t}{2} - \sinh 3H_0t\right).
\label{solution-com1}
\end{equation}
If $\psi_2 = 0$, we come back to our starting point:
a cosmological evolution (\ref{Hubblecom}), (\ref{radiuscom}) for $w = 0$. 

The case when both the coefficients $\psi_1$ and $\psi_2$ are different from zero is rather cumbersome, while its qualitative 
behavior can be understood  studying the 
case $\psi_1=0, \psi_2 \neq 0$.  
  
Let us rewrite the solution (\ref{solution-com1} for  this case in  the form
\begin{equation}
\psi_(t) = \psi_2 \frac{\cosh\frac{3H_0t}{2}} 
{\sinh\frac{3H_0t}{2}}(2-\cosh 3H_0t).
\label{solution-com2}
\end{equation}
Let us consider first the case $\psi_2 > 0$. Then,
the volume function $\psi(t)$ will be nonnegative 
in the intervals:
\begin{equation}
0 \leq t \leq \frac{1}{3H_0}arccosh 2,
\label{interval}
\end{equation}
and 
\begin{equation}
-\infty  \leq t \leq -\frac{1}{3H_0}arccosh 2.
\label{interval1}
\end{equation}
The Hubble variable can be expressed as 
\begin{equation}
h(t) = -\frac{3H_0(1-y+y^2)sign(t)}{\sqrt{y^2-1}(2-y)},
\label{Hubble-com1}
\end{equation}
and its time derivative reads
\begin{equation}
\dot{h}(t) =-\frac{9H_0^2(y^3+3y^2-6y+1)}{(y^2-1)(2-y)^2},
\label{Hubble-dot-com}
\end{equation}
where 
\begin{equation}
y \equiv \cosh 3H_0t.
\label{y-define}
\end{equation} 
One can see by  simple algebraic treatment that 
the function $\dot{h}(t)$ can have the only zero at some moment $t_0$ in the interval (\ref{interval}), 
i.e. when $0 < y < 2$.  Such a point describes a transition from a phantom-type contraction to a non-phantom contraction ending in a Big Crunch. 
The evolution in the interval (\ref{interval1}) describes instead a non-phantom contraction 
beginning at $t = -\infty$ in a de Sitter stage and ending at $t = -\frac{1}{3H_0}arccosh 2$ in the Big Crunch. 

Analogously, one can show that in the case $\psi_2 
< 0$, one can have either non-phantom expansion beginning from the Big Bang and ending in an infinite de Sitter universe or  the transition from the non-phantom expansion to phantom expansion culminating in a Big Rip singularity at 
$t = \frac{1}{3H_0}arccosh 2$. The inclusion of the term with $\psi_1 \neq 0$ can change  some features of the  cosmological evolution. However, the qualitative and numerical analysis of the corresponding differential equations shows that the number of the moments of time when the universe crosses the phantom divide line does not change.   

For example, for the case when both the coefficients $\psi_1$ and $\psi_2$ are positive and $\psi_1 = 3\psi_2$ the evolution is presented in Fig. 1.  
\begin{figure}[h]
\epsfxsize 6cm 
\epsfbox{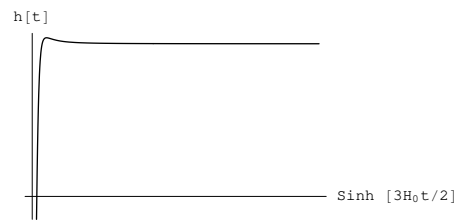}
\caption{$h(t)$ dependence at $\psi_1/\psi_2 =3$}
\label{Fig.1}
\end{figure}  
Here, the universe begins its evolution at $t = 0$
in the phase of phantom-type contraction, which at some moment (when $h(t)$ crosses the line $h=0$) is substituted by the phase of phantom-type expansion. After some period, the universe  crosses the phantom divide line (when $h(t)$ has its maximum value) and enters into the phase of an infinite normal  
expansion.

As we have already mentioned above, the number of the scalar field potentials treated as functions of the scalar field, for which it is possible to find an exact solution is very limited. Instead, one can find some
potentials treated as functions of the cosmic time 
parameter, for which one can find complete exact solutions. It is connected with the linearity of the corresponding equation (\ref{equation}), which is nothing but a 
Schr\"{o}dinger equation for a system with zero energy.  Thus, we can borrow some solvable examples of such potentials from non-relativistic quantum mechanics (see e.g. \cite{Newton,Chadan,Cooper}).
We shall give here a couple of examples of such potentials. For the potential
\begin{equation}
V(t) = \frac{A^2}{t^4}
\label{example}
\end{equation}
the solution of Eq. (\ref{equation}) looks like
\begin{equation}
\psi = \psi_t\exp\left(\frac{A}
{t}\right) + \psi_2 t\exp\left(-\frac{A}
{t}\right).
\label{example1}
\end{equation}
For the potential 
\begin{equation}
V(t) = \frac{B^2 + 1}{(1+t^2)^2}
\label{example2}
\end{equation}
the solution reads
\begin{eqnarray}
&&\psi = \psi_1\sqrt{1+t^2}\exp(B {\rm arctan} t)  
\nonumber \\
&&+\psi_2\sqrt{1+t^2}\exp(-B {\rm arctan} t).
\label{example3}
\end{eqnarray}
The investigation of cosmologies based on the solutions (\ref{example1}) and (\ref{example3})
might represent some interest but it is outside of the scope of the present paper.


\begin{thebibliography}{99}
 \bibitem{cosmic}
A. Riess et al., 
Astron. J. {\bf 116}, 1009 (1998); 
S.J. Perlmutter et al., 
Astroph. J. {\bf 517}, 565 (1999).
\bibitem{dark}
V. Sahni and A.A. Starobinsky, Int. J. Mod. Phys. D {\bf 9}, 373 (2000); 
T. Padmananbhan, 
Phys. Rep. {\bf 380}, 235 (2003);
P.J.E. Peebles and B. Ratra,
Rev. Mod. Phys. {\bf 75}, 559 (2003);
V. Sahni,
Class. Quantum Grav. {\bf 19}, 3435 (2002).
\bibitem{quintessence}
R.R. Caldwell, R. Dave and P.J. Steinhardt,
Phys. Rev. Lett. {\bf 80}, 1582 (1998);
J.A. Frieman, I. Waga, Phys. Rev. D {\bf 57}, 4642 (1998).
\bibitem{darkmodel}
C. Armendariz-Picon, T. Damour and  V. Mukhanov, Phys. Lett. B
{\bf 458}, 209 (1999); C. Armendariz-Picon, V. Mukhanov and P.J.
Steinhardt, Phys. Rev. Lett. {\bf 85}, 4438 (2000); C. Armendariz-Picon,
V. Mukhanov, P.J. Steinhardt, Phys. Rev. D {\bf 63}, 103510 (2001);
T. Chiba, T. Okabe, M. Yamaguchi,
Phys. Rev. D {\bf 62}, 023511 (2000);
A. Kamenshchik, U. Moschella and V. Pasquier, Phys. Lett. B {\bf 511},
265 (2001);
N. Bilic, G.B. Tupper, and R. Violeer, Phys. Lett. B {\bf 535}, 17
(2002);
J.C. Fabris, S,V.B. Gonsalves and P.E. de Souza, Gen. Rel. Grav.
{\bf 34}, 53 (2002):
M.C. Bento, O. Bertolami and A.A. Sen, Phys. Rev. D {\bf 66}, 043507
(2002);
V. Gorini, A.  Kamenshchik and U. Moschella, Phys. Rev. D {\bf 67},
063509 (2003);
A. Sen, JHEP {\bf 0204}, 048 (2002);
G.W. Gibbons, Phys. Lett. B {\bf 537}, 1 (2002); M. Fairbairn and M.H.G.
Tytgat, Phys. Lett. B {\bf 546}, 1 (2002); S. Mukohyama, Phys. Rev. D
{\bf 66}, 024009 (2002); T. Padmanabhan and T.R. Choudhury, Phys. Rev.
D {\bf 66}, 081301(R) (2002); M. Sami, P. Chingangbam, 
T. Qureshi, Phys. Rev. D {\bf 66}, 043530 (2002); G.N.
Felder, L. Kofman, A. Starobinsky, JHEP {\bf 0209}, 026 (2002); 
A. Frolov, L. Kofman and A. Starobinsky, Phys. Lett. B {\bf 545}, 8
(2002);
T. Padmanabhan, Phys. Rev. D {\bf 66}, 021301(R) (2002);
A. Feinstein, Phys. Rev. D {\bf 66}, 063511 (2002);
L.W. Abramo and F. Finelli, Phys. Lett. B {\bf 575}, 165 (2003);
J.S.
Bagla, H.K. Jassal and T. Padmanabhan, Phys. Rev. D {\bf 67}, 063504
(2003); G. Gibbons, Class. Quant. Grav. {\bf 20}, S321 (2003); M.-B. Causse, A rolling tachyon field for both
dark energy and dark halos of galaxies, astro-ph/0312206; B.C.
Paul, M. Sami, Phys. Rev. D {\bf 70}, 027301 (2004); A. Sen,
Phys. Scripta T {\bf 117},70 (2005); 
M.R. Garousi, M. Sami
and S. Tsujikawa, Phys. Rev. D {\bf 70}, 043536 (2004); G. Calcagni,
Phys. Rev. D {\bf 69}, 103508 (2004); J.M. Aguirregabiria and R. Lazkoz,
Mod. Phys. Lett. A  {\bf 19}, 927 (2004); J.M. Aguirregabiria and R.
Lazkoz, Phys. Rev. D {\bf 69}, 123502 (2004); R. Herrera, D. Pavon and
W. Zimdahl, Gen. Rel. Grav. {\bf 36}, 2161 (2004); N. Barnaby, JHEP
{\bf 0407}, 025 (2004);
W. Zimdahl, gr-qc/0505056.
 \bibitem{phantom}
R.R. Caldwell, Phys. Lett. B {\bf 545}, 23 (2002);
P. F. Gonzalez-Diaz,
Phys. Lett. B {\bf 586}, 1 (2004);
P. Singh, M. Sami and N. Dadhich,
Phys.  Rev.  D {\bf 68}, 023522 (2003); 
S. Capozziello, S. Nojiri and S.D. Odintsov,  Phys. Lett. B {\bf 632}, 597 (2006);
V.B. Johri,
Phys. Rev.  D {\bf 70}, 041303(R) (2004); 
V.K.Onemli and R.P.Woodard, Class. Quant. Grav.{\bf19}, 4607
(2000); 
S. Hannestad and E. Mortsell, Phys. Rev.
D {\bf66}, 063508 (2002); S. M. Carroll, M. Hoffman and M. Trodden,
Phys.  Rev.  D {\bf 68}, 023509 (2003); 
P.H. Frampton, hep-th/0302007; 
E. Elizalde, S. Nojiri and S.D. Odintsov,
Phys. Rev. D {\bf 70}, 043539 (2004);
P. F. Gonzalez-Diaz and
C. L. Siguenza, 
Nucl. Phys. B {\bf 697}, 363 (2004);
G.W.Gibbons, hep-th/0302199; B.McInnes, astro-ph/0210321;
L. P. Chimento and R. Lazkoz, Phys. Rev.  Lett.\  {\bf 91},
211301 (2003);
M.P. Dabrowski, T. Stachowiak and M. Szydlowski, 
Phys. Rev. D {\bf 68}, 103519 (2003);
P. Singh, M. Sami and N. Dadhich,
Phys. Rev. D {\bf 68}, 023522
(2003);
P.F. Gonzalez-Diaz, Phys. Rev.  D {\bf 69}, 063522
(2004); V.K. Onemli and R.P. Woodard,
Phys. Rev. D {\bf 70}, 107301 (2004);
M. Sami and A. Toporensky, 
Mod. Phys. Lett. A {\bf 19}, 1509 (2004);
H. Stefancic,
Eur. Phys. J. C {\bf 36}, 523 (2004);
Phys. Lett. B {\bf 586}, 5 (2004);
T. Brunier, V.K. Onemli and R.P. Woodard,
Class. Quantum  Grav.  {\bf 22}, 59 (2005);
J. Santos and J.S. Alcaniz,
Phys. Lett. B {\bf 619}, 11 (2005); 
F.C. Carvalho and A. Saa,
Phys. Rev. D {\bf 70}, 087302 (2004);
A. Melchiorri, L. Mersini, C. J. Odman and M. Trodden,
Phys. Rev. D {\bf 68}, 043509 (2003); 
J.M. Cline, S. Jeon and G.D. Moore,
Phys. Rev. D {\bf 70}, 043543 (2004); 
Z.K. Guo, Y.S. Piao, X.M. Zhang and Y.Z. Zhang,
Phys. Lett. B {\bf 608}, 177 (2005);  
B. Feng, X.L. Wang and X.M. Zhang,
Phys. Lett. B {\bf 607}, 35 (2005);
I.Y. Aref'eva, A.S. Koshelev and S.Y. Vernov, astro-ph/0412619; 
X. F. Zhang, H. Li, Y.S. Piao and X.M. Zhang,
astro-ph/0501652;
G. Calcagni, Phys. Rev. D {\bf 71}, 023511 (2005);
M. Li, B. Feng and  X. Zhang, JCAP {\bf 0512},002 (2005) ;
E. Babichev, V. Dokuchaev and  Yu. Eroshenko, 
Class. Quant. Grav. {\bf 22}, 143 (2005);
A. Anisimov, E. Babichev and  A. Vikman, 
JCAP {\bf 0506}, 006 (2005);
S. Nojiri, S.D. Odintsov and S. Tsujikawa,
Phys. Rev. D {\bf 71}, 063004 (2005);
B. Gumjudpai, T. Naskar,  M. Sami and  S. Tsujikawa,
JCAP {\bf 0506}, 007 (2005);
M. Sami, A. Toporensky,  P.V. Tretjakov and S. Tsujikawa, Phys. Lett. B 
{\bf 619}, 193 (2005);
S. Capozziello, S. Nojiri and S.D. Odintsov,  Phys. Lett. B {\bf 632}, 597 (2006).
\bibitem{rip}
R.R. Caldwell, M. Kamionkowski and  N.N. Weinberg, 
Phys. Rev. Lett. {\bf 91}, 071301 (2003); 
R. Kallosh, J. Kratochvil, A. Linde, E.V. Linder and M. Shmakova,
JCAP {\bf 0310}, 015 (2003).
\bibitem{star-rip}
A.A. Starobinsky, Grav. Cosmol. {\bf 6}, 157 (2000).
\bibitem{phant-obs}
U. Alam, V. Sahni, T.D. Saini and A.A. Starobinsky, 
Mon. Not. Roy. Astron. Soc. {\bf 354}, 275 (2004);
T. Padmanabhan and T.R.  Choudhury,
Mon. Not. Roy. Astron. Soc. {\bf 344}, 823 (2003);
T. R. Choudhury and  T. Padmanabhan, astro-ph/0311622;   Y. Wang and
P. Mukherjee,
Astrophys. J.   {\bf 606}, 654 (2004);
D. Huterer and A. Cooray, astro-ph/0404062;  
R.A. Daly and S. G. Djorgovski,
Astrophys.  J. {\bf 597}, 9 (2003); J.
S. Alcaniz, Phys. Rev. D {\bf69}, 083521 (2004);
J.A.S. Lima, J.V. Cunha and J.S. Alcaniz,
Phys. Rev. D {\bf 68}, 023510 (2003);
U. Alam, V. Sahni and A.A. Starobinsky, 
JCAP {\bf 0406}, 008 (2004);
Y. Wang and K. Freese,
Phys. Lett. B {\bf 632}, 449 (2006);
A. Upadhye, M. Ishak and P. J. Steinhardt,
Phys. Rev. D {\bf 72} ,063501 (2005);
D.A. Dicus and W. W. Repko,
Phys. Rev.  D {\bf 70}, 083527 (2004);
C. Espana-Bonet and P. Ruiz-Lapuente,
hep-ph/0503210;
Y. H. Wei,
astro-ph/0405368;
H.K. Jassal, J.S. Bagla, and T. Padmanabhan, 
 Mon. Not. Roy. Astron. Soc. Letters, {\bf 356}, L11 (2005);
S. Nesseris and L. Perivolaropoulos, Phys. Rev. D {\bf 70},
043531 (2004);
R. Lazkoz, S. Nesseris and L. Perivolaropoulos,
JCAP {\bf 0511}, 010 (2005) ; 
H.K. Jassal, J.S. Bagla and T. Padmanabhan,
Phys. Rev. D {\bf 72},103503 (2005); 
B. Feng, M. Li, Y.S. Piao and X. Zhang,
astro-ph/0407432;
J.-Q. Xia, B. Feng and X.-M. Zhang, 
Mod. Phys. Lett. A {\bf 20}, 2409 (2005). 
\bibitem{phant-obs1} 
A.Cabre, E.Gaztanaga, M.Manera, P.Fosalba, F.Castander,astro-ph/0603690.
\bibitem{divide}
B. Boisseau, G. Esposito-Farese, D. Polarski and A.A. Starobinsky,
Phys. Rev. Lett. {\bf 85}, 2236 (2000);
G. Esposito-Farese and D. Polarski, Phys. Rev. D {\bf 63}, 063504 (2001);
A. Vikman,
Phys.  Rev.  D {\bf 71}, 023515 (2005); 
L. Perivolaropoulos,
Phys.  Rev. D {\bf 71}, 063503 (2005);
B. McInnes, Nucl. Phys. B {\bf 718}, 55 (2005);
I.Ya. Aref'eva, A.S. Koshelev and S.Yu. Vernov,  Phys. Rev. D {\bf 72}, 064017 (2005);
L. Perivolaroupoulos,  JCAP {\bf 0510}, 001 (2005);
R.R. Caldwell and M. Doran,
Phys. Rev. D {\bf 72}, 043527 (2005);
 A.A. Andrianov, F. Cannata and A.Y. Kamenshchik,
 gr-qc/0512038;
H. Wei, R.-G. Cai and D.-F. Zeng, Class. Quant. Grav. {\bf 22}, 3189 (2005);
H. Wei, R.-G. Cai, Phys. Rev. D {\bf 72}, 123507(2005) ; astro-ph/0512018.
 \bibitem{we}
 A.A. Andrianov, F. Cannata and A.Y. Kamenshchik,
 Phys. Rev. D {\bf 72}, 043531 (2005).
 \bibitem{exp}
F. Lucchin and S. Matarrese, Phys. Rev. D {\bf 32}, 1316 (1985);
J.J. Halliwell, Phys. Lett. B {\bf 185}, 341 (1987);
J.D. Barrow, Phys. Lett. B {\bf 187}, 12 (1987);
A.B. Burd and J.D. Barrow, Nucl. Phys. B {\bf 308}, 929 (1988);
B. Ratra, Phys. Rev. D {\bf 45}, 1913 (1992);
J.D. Barrow, Phys. Lett. B {\bf 235}, 40 (1990);
J.E. Lidsey, Class. Quantum Grav. {\bf 9}, 1239 (1992);
S. Capozziello, R. de Ritis, C. Rubano and P. Scudellaro, Riv. 
Nuovo Cimento {\bf 19}, 1 (1996); A.A. Coley, J. Ibanez and 
R.J. van den Hoogen, J. Math. Phys. {\bf 38}, 5256 (1997);
C. Rubano and P. Scudellaro, Gen. Rel. Grav. {\bf 34}, 307 (2002).  
\bibitem{Chervon}
S.V. Chervon and V.M. Zhuravlev, gr-qc/9907051.
\bibitem{Yurov}
A.V. Yurov, astro-ph/0305019. 
\bibitem{Yurov1}
A.V. Yurov and S.D. Vereshchagin, Theor. Math. Phys. {\bf 139}, 787 (2004).
\bibitem{reconstruct}
J. Simon, L. Verde and R. Jimenez, Phys. Rev. D {\bf 71}, 123001 (2005);
Z.-K. Guo, N. Ohta and Y.-Z. Zhang, Phys. Rev. D {\bf 72}, 023504 (2005);
astro-ph/0603109.
\bibitem{we-tach}
V. Gorini, A. Kamenshchik, U. Moschella and V. Pasquier, Phys. Rev. D {\bf 69}, 123512 (2004).
 \bibitem{Newton}
 R.G. Newton, {\it Scattering theory of waves and particles} (McGraw-Hill, New York, 1966). 
 \bibitem{Chadan}
 K. Chadan and R. Kobayashi, math-ph/0510047;
 math-ph/0602002.
\bibitem{Cooper}
F. Cooper, A. Khare and U.P. Sukhatme, {\it Supersymmetry in Quantum Mechanics} (World  Sceintific,
Singapore, 2001).   
\end{thebibliography}
 \end{document}